\documentclass{moriond}

\usepackage[symbol]{footmisc}
\usepackage[english]{babel}
\usepackage{graphicx}
\usepackage{graphics}
\usepackage{braket}
\usepackage{bbold}
\usepackage{amssymb}
\usepackage{amsmath}
\usepackage{nicefrac}
\usepackage{dcolumn}% Align table columns on decimal point
\usepackage{bm}% bold math
\usepackage{slashed}
\usepackage{datetime}
\usepackage{mciteplus}
\usepackage{multirow}
\usepackage{bbold}
\usepackage{siunitx}
\usepackage{booktabs}
\usepackage{color, soul}
\usepackage[usenames,dvipsnames]{xcolor}
\usepackage{float}
\usepackage[utf8]{inputenc}
\usepackage[normalem]{ulem}
\usepackage{mathtools}

\newcommand{\LQCD}{\Lambda_{\rm QCD}}

\def\be{\begin{equation}}
\def\ee{\end{equation}}
\def\bea{\begin{eqnarray}}
\def\eea{\end{eqnarray}}

\newcommand{\Photo}{\includegraphics[height=45mm]{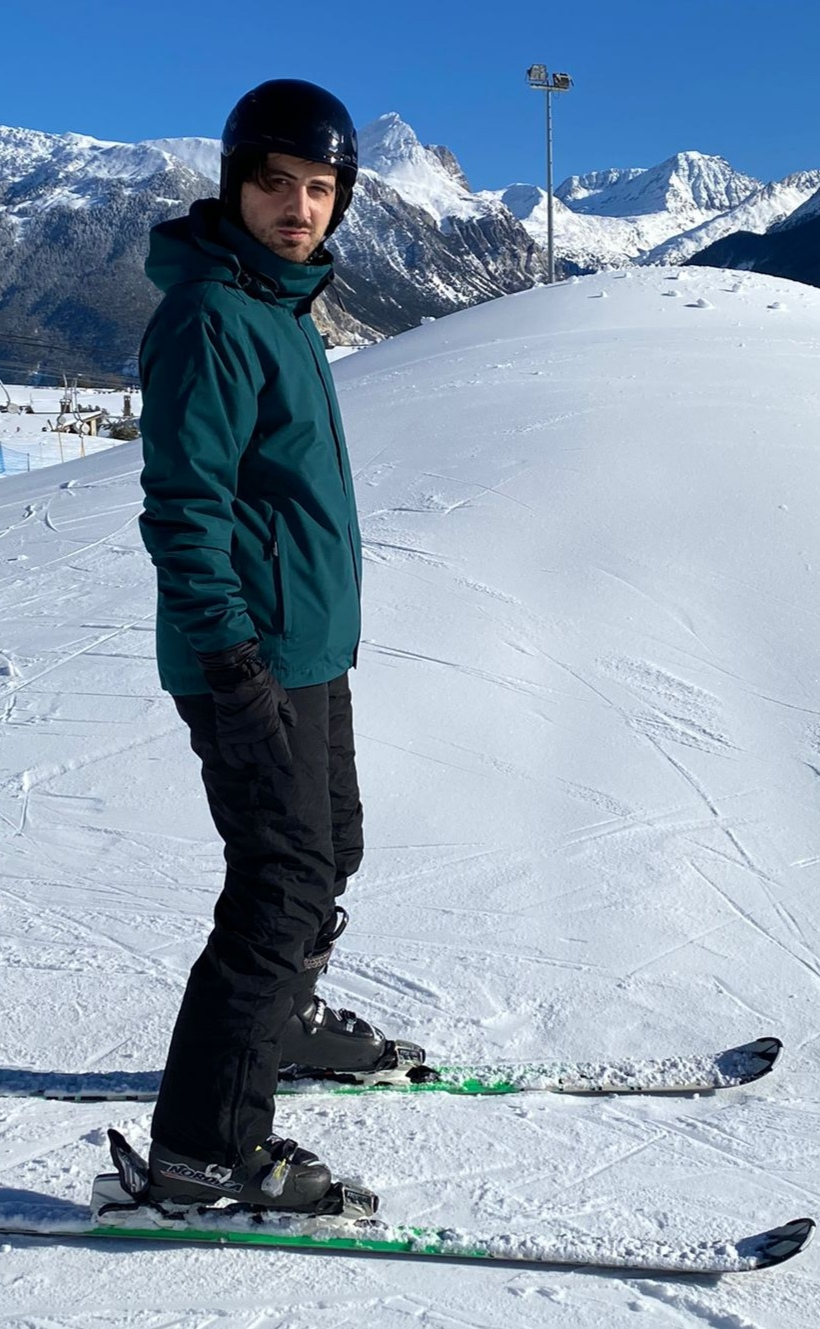}}

\begin{document}
\vspace*{4cm}
\title{Higgs boson production at next-to-leading logarithmic accuracy}

\author{F.G.~Celiberto,$^{\;1,2\;}$ M.~Fucilla,$^{\;3,4,5}$~\footnote[2]{Corresponding Author. Electronic address: \href{mailto:michael.fucilla@unical.it}{michael.fucilla@unical.it}} D.Yu.~Ivanov,$^{\;6}$ M.M.A.~Mohammed$^{\;3,4}$ and A.~Papa$^{\;3,4}$}

\address{
${}^1$INFN-TIFPA Trento Institute of Fundamental Physics and Applications, I-38123 Povo, Trento, Italy
\\
${}^2$Universidad de Alcal\'a (UAH), E-28805 Alcal\'a de Henares, Madrid, Spain
\\
${}^3$Dipartimento di Fisica, Universit\`a della Calabria, I-87036 Arcavacata di Rende, Cosenza, Italy
\\
${}^4$INFN-Cosenza, I-87036 Arcavacata di Rende, Cosenza, Italy
\\
${}^5$Universit\'e Paris-Saclay, CNRS, IJCLab, 91405 Orsay, France
\\
${}^6$Sobolev Institute of Mathematics, 630090 Novosibirsk, Russia \vspace{0.2 cm}
} 

\maketitle\abstracts{
%The kinematics reachable at modern and future colliders offers us a unique opportunity to probe \textit{Quantum Chromodynamics} in quite original and unprecedented ways. In the so-called \textit{Regge–Gribov} limit, also known as \textit{semi-hard} regime [* non sono sicuro che "Regge-Gribov limit" o "semi-hard" siano sinonimi*], the presence of the strong scale hierarchy $\sqrt{s} \gg Q $, where $s$ is the center-of-mass energy and $ Q $ the typical hard scale of the process, leads to the growth of logarithmic contributions of the form $\ln s/ Q^2$, which have to be resummed to all orders.
It has been recently argued that the inclusive hadroproduction at the LHC of a Higgs boson in association with a jet can be sensitive to the high-energy dynamics. Moreover, the impact of the resummation at FCC energies is expected to be large also in the inclusive cross section for the main Higgs production channel in proton-proton collisions, namely the gluon fusion. 
As the energy increases, a pure $k_t$-factorization framework or a collinear-factorization approach supplemented by the high-energy resummation are adequate formalisms to describe these processes. In both cases, the fundamental missing ingredient for a next-to-leading logarithmic description is the Higgs boson impact factor. We present the full next-to-leading result for the forward Higgs impact factor, obtained in the infinite top-mass limit, discussing possible future extensions, such as the calculation of the impact factor in the central rapidity region.}

\section{Introduction}
\label{sec:introduction}
Energy logarithms arise in the so-called Regge--Gribov limit, accessed via LHC \emph{semi-hard} processes. Here, the $\LQCD \ll \{Q\} \ll \sqrt{s}$ strong scale hierarchy, with $\LQCD$ the QCD hadronization scale, $\{Q \}$ one or a set of process-dependent hard scales, and $s$ the squared center-of-mass energy, leads to the growth of logarithmic contributions of the form $\ln (s / Q^2)$, which have to be resummed to all orders. The established tool for performing this resummation is the Balitsky–Fadin–Kuraev–Lipatov (BFKL) approach,~\cite{Fadin:1975cb,Balitsky:1978ic} which holds within the LLA (leading logarithmic approximation), in which all terms $(\alpha_s \ln s)^n$ are resumed, and in the NLLA (next-to-leading logarithmic approximation), in which also all terms $\alpha_s (\alpha_s \ln s)^n$ are included (see~\cite{Ducloue:2013bva}$^-$\cite{Celiberto:2022dyf} for applications). In the BFKL approach, a generic scattering amplitude can be expressed as the convolution of a process-independent Green’s function with two impact factors, related to the transition from each colliding particle to the final-state object in the respective fragmentation region. The BFKL Green’s function satisfies an integral equation, whose kernel is known at the NLO for any fixed (not growing with $s$) momentum transfer, $t$, and for any possible two-gluon color
configuration in the $t$-channel. However a full NLA realization of the BFKL approach requires the knowledge of NLO impact factors. We refer to~\cite{Caucal:2021ent}$^-$\cite{Taels:2022tza} for recent progress in the computation of NLO impact factors.
Here, we discuss a new ingredient, the forward Higgs boson impact factor.~\cite{Celiberto:2022fgx} It was independently computed in the Lipatov effective action.~\cite{Nefedov:2019mrg,Hentschinski:2020tbi}  

\section{BFKL factorization and LO impact factor}
The proton-initiated impact factor is related to the gluon-initiated one by the factorization
 \begin{equation*}
    \frac{d \Phi_{PP}^{H}}{d x_H d^2 \vec{p}_H} = \int_{x_H}^1 \frac{d z_H}{z_H} f_g \left( \frac{x_H}{z_H} \right) \frac{d \Phi_{gg}^{H}}{d z_H d^2 \vec{p}_H} \; ,
 \end{equation*}
where $\vec{p}_H$ is the transverse momentum of the Higgs with respect to the collision axis, $z_H$ ($x_H$) is its fraction momenta in the direction of the gluon (proton), and $f_g$ is the gluon parton distribution function (PDF).
At LO, the forward-Higgs impact factor reads~\cite{Celiberto:2020tmb}
\label{sec:LOforwardHiggs}
 \begin{equation}
          \frac{d \Phi_{PP}^{\{ H \}(0)}}{d x_H d^2 \vec{p}_H} = \frac{\alpha_s^2}{v^2} \frac{ \vec{q}^{\; 2}|\mathcal{F}(m_t, m_H, \vec{q}^{\; 2})|^2}{128 \pi^2 \sqrt{2(N^2-1)}} f_g (x_H) \delta^{(2)} (\vec{p}_H -\vec{q}) \; ,
\end{equation}
where $\vec{q}$ is the momentum of the Reggeized $t$-channel gluon, $v$ is the electroweak vacuum expectation value parameter, $v = 1/(G_F \sqrt{2})$, and $m_{t,H}$ the top-quark and Higgs masses. The expression of $\mathcal{F}$ can be found in.~\cite{Celiberto:2020tmb}
In the infinite top-mass limit, the impact factor reduces to 
\begin{equation}
    \frac{d \Phi_{PP}^{ \{ H \}(0)}}{d x_H d^2 \vec{p}_H} = \frac{g_H^2 \vec{q}^{\; 2} f_g (x_H) \delta^{(2)} ( \vec{q} - \vec{p}_H) }{8 \sqrt{N^2-1}} \; , \hspace{0.5 cm} g_H = \frac{\alpha_s}{3 \pi v} \left( 1 + \frac{11}{4} \frac{\alpha_s}{\pi} \right) + {\cal O} (\alpha_s^3) \; .
\end{equation}
This result could be obtained by making use of the effective Lagrangian
\begin{equation}
\mathcal{L}_{ggH} = - \frac{g_H}{4} F_{\mu \nu}^{a} F^{\mu \nu,a} H \; ,
\label{EffLagrangia}
\end{equation}
which couples the Higgs field to gluons directly, via the QCD field strength tensor $F_{\mu \nu}^a = \partial_{\mu} A_{\nu}^a - \partial_{\nu} A_{\mu}^a  + g f^{abc} A_{\mu}^b A_{\nu}^c \;$. In the next sections, we employ the effective Lagrangian in Eq.~(\ref{EffLagrangia}) to compute the NLO corrections to the Higgs impact factor in the infinite top-mass limit.   

\section{Real corrections}
\label{sec:RealCorrections}
Beyond LO, the hard sub-process for the production of the Higgs boson can be initiated either by a quark or by a gluon. The quark-initiated contribution to the impact factor reads
\begin{equation}
    \frac{d \Phi_{q q}^{\{H q \}} (z_H, \vec{p}_H, \vec{q})}{d z_H d^2 \vec{p}_H} = \frac{\sqrt{N^2-1}}{16 N (2 \pi)^{D-1}} \frac{g^2 g_H^2}{(\vec{r}^{\; 2})^2}
   \left[ \frac{4 (1-z_H) \left( \vec{r} \cdot \vec{q} \; \right)^2 + z_H^2 \vec{q}^{\; 2} \vec{r}^{\; 2}}{z_H} \right] \; ,
    \label{QuarkConImpacFin}
\end{equation} 
while the gluon-initiated contribution reads
\begin{gather}
\frac{d \Phi_{g g}^{\{H g \}} (\vec{q} \; )}{d z_H d^{2} \vec{p}_H} = \frac{g^2 g_H^2 C_A}{8 (2 \pi)^{D-1}(1-\epsilon) \sqrt{N^2-1}} \frac{2 \vec{q}^{\; 2}}{\vec{r}^{\; 2}} \nonumber \\
      \times \left[ \frac{z_H}{1-z_H} + z_H (1-z_H) + 2 (1-\epsilon) \frac{(1-z_H)}{z_H} \frac{(\vec{q} \cdot \vec{r})^2}{\vec{q}^{\; 2} \vec{r}^{\; 2}} \right]  \theta \left( s_{\Lambda} - s_{gR} \right) + \rm{finite \ terms} \; ,
     \label{GluonImp}
\end{gather}
where $s_{gR}$ is the gluon-Reggeon squared invariant mass. 
We explicitly show only the impact-factor part featuring phase-space singularities.
Both the contributions have a \textit{collinear} singularity when $\vec{r} = \vec{q}-\vec{p}_H \rightarrow \vec{0}$. In addition, the gluon contribution features \textit{rapidity} divergences, when $z_H \rightarrow 1$, and \textit{soft} divergences, when both $\vec{r} \rightarrow \vec{0}$ and $z_H \rightarrow 1$. The rapidity divergence is regularized by the $\theta (s_{\Lambda}-s_{gR})$, which prevents the Higgs and the additional emitted gluon to be strongly separated in the rapidity. It is then removed by a suitable ``BFKL counterterm".~\cite{Celiberto:2022fgx} 

\section{Virtual corrections}
\label{sec:VirtualCorrections}
Being impact factors obtained by squaring effective vertices, we must extract the $1$-loop effective vertex for the production of a Higgs in gluon-Reggeon collisions. To this aim, we use a reference amplitude and compare it with the expected Regge form. We employ the amplitude for the diffusion of a gluon off a quark to produce a Higgs plus a quark, $\mathcal{A}_{gq \rightarrow Hq}^{(8,-)}$, with octet color state and negative signature in the $t$-channel. It should assume the following Reggeized form~\footnote{The apexes $(0)$ and $(1)$ 
denote the Born and the $1$-loop approximation, respectively.}
$$
{\cal A}_{g q \rightarrow H q}^{(8,-)} = \Gamma_{ H g}^{ac} \frac{s}{t}\left[ \left( \frac{s}{-t} \right)^{\omega(t)} + \left(
\frac{-s}{-t} \right)^{\omega(t)} \right]\Gamma_{qq}^{c} \approx
\Gamma_{H g}^{ac(0)} \frac{2s}{t} \Gamma_{qq}^{c(0)}
$$
\begin{equation}
+ \Gamma_{H g}^{ac(0)} \frac{s}{t}\omega^{(1)}(t)
\left[ \ln\left( \frac{s}{-t} \right) + \ln\left(
\frac{-s}{-t} \right) \right]\Gamma_{q q}^{c(0)} + \Gamma_{H g}^{ac(0)} \frac{2s}{t}\Gamma_{q q}^{c(1)} +
\Gamma_{H g}^{ac(1)}\frac{2s}{t} \Gamma_{qq}^{c(0)} \; ,
\label{ReggeFormEx1}
\end{equation}
where $\omega (t)$ is the Regge trajectory, $ \Gamma_{ H g}^{ac}$ is the LO gluon-Higgs-Reggeon effective vertex and lastly $\Gamma_{qq}^{c}$ is the quark-quark-Reggeon effective vertex.  Since the only unknown ingredients in the right-hand side of Eq.~(\ref{ReggeFormEx1}) is the one-loop correction to the gluon-Higgs-Reggeon vertex, if we compute the amplitude ${\cal A}_{g q \rightarrow H q}^{(8,-)}$, we are immediately able to extract it. The effective vertex allows us to obtain the virtual contribution to the impact factor, which reads
\begin{equation*}
\frac{d \Phi_{gg}^{\{ H \}(1)}}{d z_H d^2 \vec{p}_H} = \frac{d \Phi_{gg}^{\{ H \}(0)}}{d z_H d^2 \vec{p}_H} \; \frac{\bar{\alpha}_s}{2 \pi} \left( \frac{\vec{q}^{\; 2}}{\mu^2}  \right)^{- \epsilon}  \left[  - \frac{C_A}{\epsilon^2} + \frac{11 C_A - 2n_f}{6 \epsilon} \right.
\end{equation*}
\begin{equation}
      \left. - \frac{C_A}{\epsilon} \ln \left( \frac{\vec{q}^{\; 2}}{s_0} \right) - \frac{5 n_f}{9} + C_A \left( 2\;\Re \left( {\rm{Li}}_2 \left( 1 + \frac{m_H^2}{\vec{q}^{\; 2}} \right)\right) + \frac{\pi^2}{3} + \frac{67}{18} \right) + 11 \right] \; .
      \label{VirtualPartIMF}
\end{equation}
In the computation $\epsilon=\epsilon_{UV}=\epsilon_{IR}$ is set such that any scaleless Feynman integral does not contribute to virtual corrections.
%\begin{figure}
 %   \centering
  %  \includegraphics[scale=0.3]{PrimaCorr-eps-converted-to.pdf} \hspace{0.5 cm}
  %  \includegraphics[scale=0.3]{SecondaCorr-eps-converted-to.pdf} \hspace{0.5 cm}
  %  \includegraphics[scale=0.3]{TerzaCorr-eps-converted-to.pdf} \vspace{0.5 cm} \\
   % \includegraphics[scale=0.3]{QuartaCorr-eps-converted-to.pdf} \hspace{0.5 cm}
   % \includegraphics[scale=0.3]{QuintaCorr-eps-converted-to.pdf} \hspace{0.5 cm}
   % \includegraphics[scale=0.3]{SestaCorr-eps-converted-to.pdf} \vspace{0.3 cm} \\
   % \includegraphics[scale=0.3]{SettimaCorr-eps-converted-to.pdf} \hspace{0.3 cm}
   % \includegraphics[scale=0.3]{OttavaCorr-eps-converted-to.pdf} \hspace{0.3 cm}
   % \includegraphics[scale=0.3]{NonaCorr-eps-converted-to.pdf}
   % \label{VirtualCorrections}
   % \caption{Diagrams contributing to the next-to-leading order virtual corrections of the $\mathcal{A}_{gq \rightarrow Hq}^{(8,-)}$ amplitude}
%\end{figure}

\section{Projection onto the eigenfunctions of the LO BFKL kernel}
In the previous section we gave the transverse momentum representation of the impact factor. Then, the projection of the impact factor onto the eigenfunctions of the LO BFKL kernel,
\begin{equation}
   \frac{d \Phi_{A A}(n,\nu)}{d x_H d^2 \vec{p}_H}  \equiv \int \frac{d^{2-2\epsilon} q}{\pi \sqrt{2}} (\vec{q}^{\; 2})^{i \nu - \frac{3}{2}} e^{i n \phi} \frac{d \Phi_{A A} (\vec{q} \;)}{d x_H d^2 \vec{p}_H} \;,
\label{Projection}
\end{equation}
is very useful to observe the cancellation of divergences and for some practical applications.
Performing the convolution in Eq.~(\ref{Projection}), the phase-space singularities appearing in the real part of the Higgs impact factor can be extracted in the form of poles in dimensional regularization. In addition, we must perform the renormalization of $\alpha_s$ and $f_g(x, \mu)$. Lastly, all singularities are removed by summing real and virtual contributions, so that a finite result is obtained. The final expression can be written in terms of Gauss hypergeometric functions.~\cite{Celiberto:2022fgx}

\section{Conclusions and outlook}
\label{sec:Conclusions}
The impact factor presented here can be used to describe the inclusive hadroproduction of a forward Higgs in the limit of small Bjorken $x$, as well as for interesting studies on inclusive forward emissions of a Higgs boson in association with a backward identified object. The latter has been proposed and investigated within a partial NLLA (NLLA Green function plus LO impact factors).~\cite{Celiberto:2022fgx} In the LLA, it has also been studied in the {\tt High Energy Jets} ({\tt HEJ}) framework.~\cite{Andersen:2022zte} Studies in the original BFKL framework can now be extended to the full NLLA, and a very intriguing extension would be carrying out a matching procedure to a full NLO fixed-order calculation obtained through the {\tt POWHEG} method.~\cite{Nason:2004rx}$^-$\cite{Celiberto:2023uuk} Another fascinating perspective is the calculation of the Higgs impact factor via gluon fusion in the central-rapidity region.

\section*{Acknowledgments}

We thank Maxim Nefedov, Samuel Wallon, Lech Szymanowski and Saad Nabeebaccus for insightful discussions. This  work  is  supported  by  the  Atracci\'on  de  Talento  Grant  n.   2022-T1/TIC-24176, Madrid, Spain, and by the INFN/QFT@COLLIDERS Project, Italy.

%\section*{Appendix}

\end{document}